\documentclass[letterpaper,english,aps,showpacs,showkeys]{revtex4-1}
\usepackage[T1]{fontenc}
\usepackage[latin9]{inputenc}
\usepackage{color}
\usepackage{textcomp}
\usepackage{amsmath}
\usepackage{amssymb}
\usepackage{graphicx}
\usepackage{esint}

\makeatletter

\@ifundefined{showcaptionsetup}{}{%
 \PassOptionsToPackage{caption=false}{subfig}}
\usepackage{subfig}
\makeatother

\usepackage{babel}
\begin{document}

\title{The Effects of the modified scalar product on some properties of
the one-dimensional harmonic oscillator with energy-dependent potential}

\author{Abdelmalek Boumali}
\email{boumali.abdelmalek@gmail.com}

\selectlanguage{english}%

\affiliation{Laboratoire de Physique Appliquée et Théorique, ~\\
University Larbi Tébessi -Tébessa-, 12000, W. Tébessa, Algeria.}

\author{Samia Dilmi}
\email{sam1dilmi@yahoo.fr}

\selectlanguage{english}%

\affiliation{VTRS et Institut des Sciences et technologie, Centre universitaire
d'El-Oued., ~\\
BP.789, El-Oued 39 000, Algeria}

\author{Hassan Hassanabadi}
\email{h.hasanabadi@shahroodu.ac.ir}

\selectlanguage{english}%

\affiliation{\textcolor{black}{Physics Department, University of Shahrood, Shahrood,
Iran.}}

\author{Soroush Zare }
\email{soroushzrg@gmail.com}

\selectlanguage{english}%

\affiliation{\textcolor{black}{Physics Department, University of Shahrood, Shahrood,
Iran.}}

\date{\today}
\begin{abstract}
In this article, we try to test the influence of the modification
of the scalar product, found in the problems of the energy-dependent
potential, on the physical properties of the harmonic oscillator in
one dimension. For this, we at first discuss the effect of this change
on the thermodynamic properties of this oscillator, and then on the
parameters of Fisher and Shannon of quantum information. For the second
problem, we are an obligation to redefine this parameters. Finally,
the uncertainly relation of Cramer-Rao is well recovered in our problem
in question.
\end{abstract}

\pacs{03.65.Ge; 03.67.-a; 03.65.Ta}

\keywords{Fisher parameter; Shannon entropy; partition function; Cramer-Rao
uncertainly relation}
\maketitle

\section{Introduction}

Wave equations with energy dependent potentials have been come to
view for long time. They can be visible in Klein-Gordon equation considering
particle in an external electromagnetic field\citep{1}. Arising from
momentum dependent interactions, they also can be appear in non-relativistic
quantum mechanics, as shown by Green \citep{2} for instance Pauli-Schrodinger
equation possess another example \citep{3,4}. Sazdjian \citep{5}
and Formanek et al \citep{6} have noted that the density probability,
or the scalar product, has to be modified with respect to the usual
definition, in order to have a conserved norm. Garcia-Martinez et
al and Lombard made an investigation on Schrödinger equation with
energy-dependent potentials by solving them exactly in one and three
dimensional \citep{7,8} as well as Hassanabadi et al studied D-dimensional
Schrodinger equation for an energy-dependent Hamiltonian that linearly
depends on energy and quadratic on the relative distance \citep{9}
and also they studied in another case Dirac equation for an energy-dependent
potential in the presence of spin and pseudospin symmetries with arbitrary
spin-orbit quantum number . We calculate the corresponding eigenfunctions
and eigenvalues of this system by comparing with analytically solvable
energy-dependent potentials \citep{10}. As an example of a many-body
problem with an energy-dependent potential, we can mention what was
done by Lombard and Mareš \citep{11}. They considered systems of
N bosons bound by two-body harmonic interactions, whose frequency
depends on the total energy of the system and there are an interesting
efforts in this literature in (see Refs. \citep{12,13,14,15}).

The presence of the energy dependent potential in a wave equation
has several non-trivial implications. The most obvious one is the
modification of the scalar product, necessary to ensure the conservation
of the norm. This modification can modified some behavior or a physical
properties of a physical system: here we note that this question,
in best of our knowledge, has not been considered in the literature.
In this context, the main goal of this paper is to study the effects
of the modified scalar product of the wave function with potential
dependent energy of a typical system such as the one-dimensional Harmonic
oscillator. For this, we are focused on the study of: (i) the thermal
properties of the 1D harmonic oscillator and, (ii) the Fisher and
Shannon parameters of quantum information. This choice is justified
by the dependence of these two parameters with the density of states.
We propose in this article that in Sec. 2 harmonic oscillator with
an energy-dependent frequency has been considered and discussed by
deriving eigenfunctions, values density of a state at first. Then
the effect of modified scalar product on the thermal properties will
be the subject of the sec. III. In sec . IV, we, also, study such
effects but on the Fisher and Shannon parameters of quantum information.
Sec. V will be a conclusion.

\section{One-dimensional Harmonic oscillator with an energy-dependent potential}

\subsection{The eigensolutions}

We consider the harmonic oscillator potential with an energy-dependent
frequency
\begin{equation}
H=\frac{p^{2}}{2m}+\frac{m\omega^{2}\left(1+\gamma E^{\nu}\right)}{2}x^{2}.\label{eq:1}
\end{equation}
where $\gamma$ is a parameter (not necessarily small, though most
of our investigations are dedicated to small $\gamma$). This Hamiltonian
can be written in $\left\{ x\right\} $ representation as follows:
\begin{equation}
\hat{H}=-\frac{\hbar^{2}}{2m}\frac{d^{2}}{dx^{2}}+\frac{m\omega^{2}\left(1+\gamma E^{\nu}\right)}{2}x^{2},\label{eq:2}
\end{equation}
with in the position representation the wave function is a function
of $x$, i.e., is given by the function $\psi\left(x\right)$; the
momentum and position operators are 
\begin{equation}
p\rightarrow-i\hbar\frac{d}{dx},\,x\rightarrow\hat{x},\label{eq:3}
\end{equation}
Consequently, the time-independent Schrodinger equation for the oscillator
can be written
\begin{equation}
\left\{ -\frac{\hbar^{2}}{2m}\frac{d^{2}}{dx^{2}}+\frac{m\omega^{2}\left(1+\gamma E^{^{\nu}}\right)}{2}x^{2}\right\} \psi\left(x\right)=E\psi\left(x\right).\label{eq:4}
\end{equation}
with the following substitutions,
\begin{equation}
\xi=\sqrt{\frac{2m\omega\sqrt{1+\gamma E^{\nu}}}{\hbar}}x,\,\varepsilon=\frac{E}{\hbar\omega\sqrt{1+\gamma E^{\nu}}},\,\,\phi\left(\xi\right)=\psi\left\{ x\left(\xi\right)\right\} \label{eq:5}
\end{equation}
we obtain
\begin{equation}
\left\{ \frac{\partial^{2}}{\partial\xi^{2}}+\varepsilon-\frac{\xi^{2}}{4}\right\} \phi(\xi)=0.\label{eq:6}
\end{equation}
The last equation has the same form as
\begin{equation}
y^{''}\left(z\right)+\left\{ \vartheta+\frac{1}{2}-\frac{z^{2}}{4}\right\} \phi(z)=0,\label{eq:7}
\end{equation}
where the solutions are
\begin{equation}
y\left(z\right)=C_{1}D_{\vartheta}\left(z\right)+C_{2}D_{-\vartheta-\frac{1}{2}}\left(iz\right).\label{eq:8}
\end{equation}
Here $D_{\vartheta}\left(z\right)$ are the Weber Hermite functions,
$C_{1,2}$ are the constants and $z$ and $\nu$ are complex numbers.
In order that the solution $y(z)$ can be interpreted as wave function,
it must to check $\text{lim}_{\beta\rightarrow\pm\infty}\left|y\left(z\right)\right|^{2}=0$,
for a real argument of $z$. Only the function $D_{\nu}(z)$ has this
property if $\nu$is an integer. In that case
\begin{equation}
D_{\vartheta}\left(z\right)=e^{-\frac{z^{2}}{2}}He_{\vartheta}\left(z\right),\,\vartheta=0,1,2,\ldots\label{eq:9}
\end{equation}
where
\begin{equation}
He_{\vartheta}\left(z\right)=\left(-1\right)^{\vartheta}e^{\frac{z^{2}}{2}}\frac{d^{\vartheta}}{dz^{\vartheta}}e^{-\frac{z^{2}}{2}},\label{eq:10}
\end{equation}
are related to the Hermite polynomials $H_{\nu}\left(z\right)$ via
\begin{equation}
He_{\vartheta}\left(z\right)=\sqrt{2}^{-\vartheta}H_{\vartheta}\left(\sqrt{2}z\right).\label{eq:11}
\end{equation}
Finally, the eigensolutions are
\begin{equation}
\psi_{n}\left(x\right)=C_{n}e^{-\frac{m\omega\sqrt{1+\gamma E^{\nu}}x^{2}}{2\hbar}}H_{n}\left(\sqrt{\frac{m\omega\sqrt{1+\gamma E^{\nu}}}{\hbar}}x\right),\label{eq:12}
\end{equation}
\begin{equation}
E^{2}-\hbar^{2}\omega^{2}\left(n+\frac{1}{2}\right)^{2}\gamma E^{\nu}-\hbar^{2}\omega^{2}\left(n+\frac{1}{2}\right)^{2}=0.\label{eq:13}
\end{equation}
Now, Let us consider the following two particular cases (we use that
$\hbar=\omega=m=1$, and $\gamma<0$):
\begin{itemize}
\item First case: $\nu=1$, and $\omega^{2}=1+\gamma E$: the eigenvalues
are
\begin{equation}
E_{1n}=\frac{\gamma}{8}\left(2n+1\right)^{2}\pm\left(n+\frac{1}{2}\right)\sqrt{1+\frac{\gamma^{2}}{16}\left(2n+1\right)^{2}},\label{eq:14}
\end{equation}
\item Second case:$\nu=2$, and $\omega^{2}=1+\gamma E^{2}$: so we obtain
\begin{equation}
E_{2n}=\pm\frac{2n+1}{\sqrt{4-\gamma\left(2n+1\right)^{2}}}.\label{eq:15}
\end{equation}
\end{itemize}
The corresponding wave function in both cases is written in compact
form by
\begin{equation}
\psi_{\alpha n}\left(x\right)=C_{\alpha n}e^{-\frac{\lambda_{\alpha}x^{2}}{2}}H_{n}\left(\sqrt{\lambda_{\alpha}}x\right),\label{eq:16}
\end{equation}
with 
\begin{equation}
\lambda_{\alpha}=\sqrt{1+\gamma E_{\alpha n}^{\nu}},\,\text{where}\,\alpha=\left(1,2\right).\label{eq:17}
\end{equation}
In both cases, only the positive energies are retained, since the
negative ones are not normalizable.

Now we are in the main goal of our study,i,e., the influence of the
modified scalar product on the wave function with energy-dependent
potential: especially on the properties of eigenvalues. According
to the works of \citep{6,8,5}, the definition of the density has
to be modified in order to ensure the validity of the continuity equation.
So, for the non-relativistic Schrodinger equation, the new definition
for the density of a state ${n}$ is given by
\begin{equation}
\rho_{n}\left(x\right)=\left|\psi_{n}\left(x\right)\right|^{2}\left(1-\frac{dV}{dE}\right).\label{eq:18}
\end{equation}
In order to represent a physical system, the density has to be positive
definite, which means here
\begin{equation}
1-\frac{dV}{dE}\geq0.\label{eq:19}
\end{equation}
This imposes constraints on the energy dependence for the theory to
be coherent: by this, we mean a theory that have the following properties:
(i) the necessary modification of the definition of probability density,
(ii) The vectors corresponding to stationary states with different
energies must be orthogonal, (ii) The formulation of the closure rule
in terms of wave functions of stationary states justifies their standardization,
(iv) finally, the operators of observables are all self-adjoint (Hermitian)\citep{8,12}.

\begin{figure}
\subfloat[first case: $\nu=1$]{\includegraphics[scale=0.3]{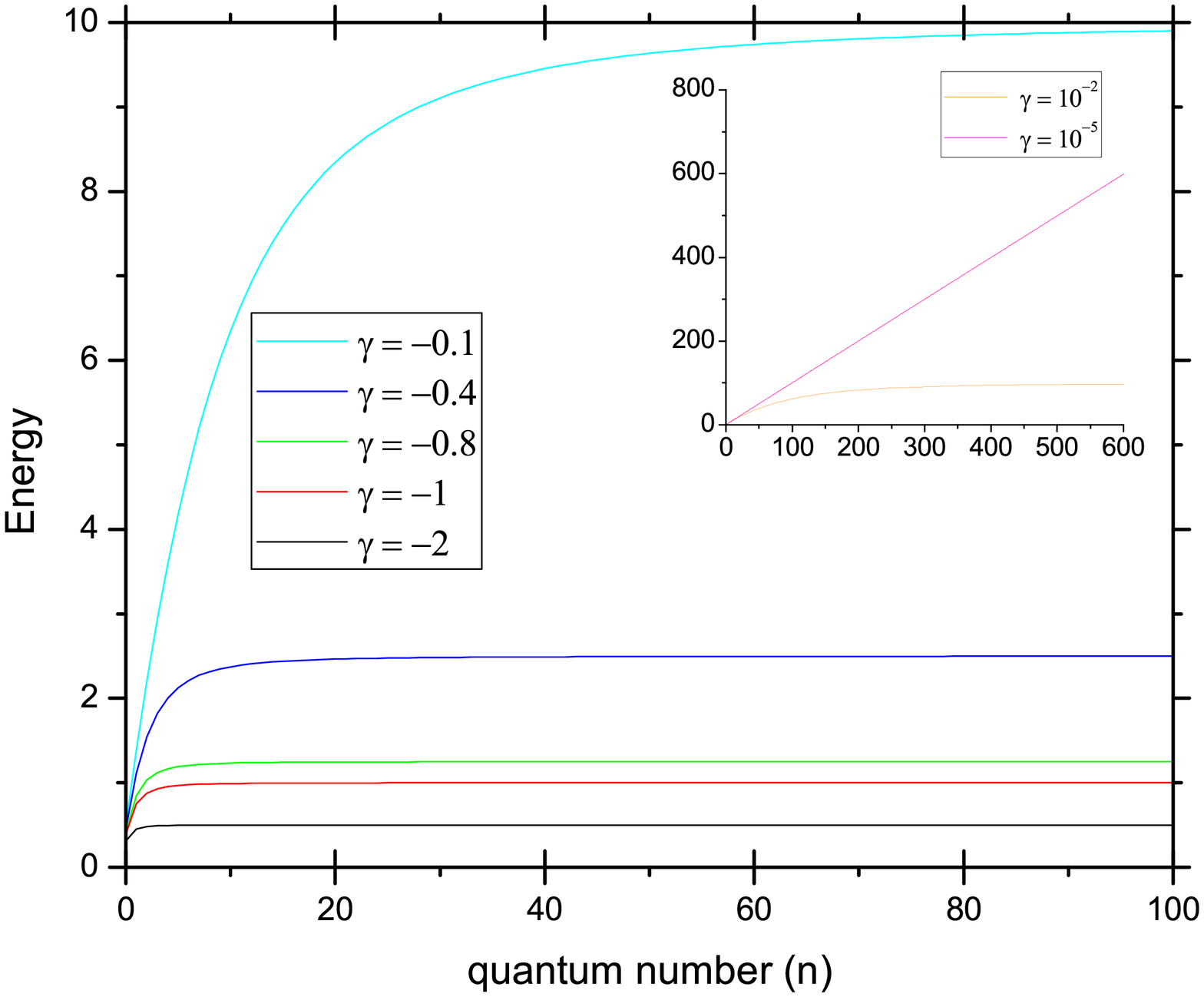}

} \subfloat[Second case: $\nu=2$]{\includegraphics[scale=0.3]{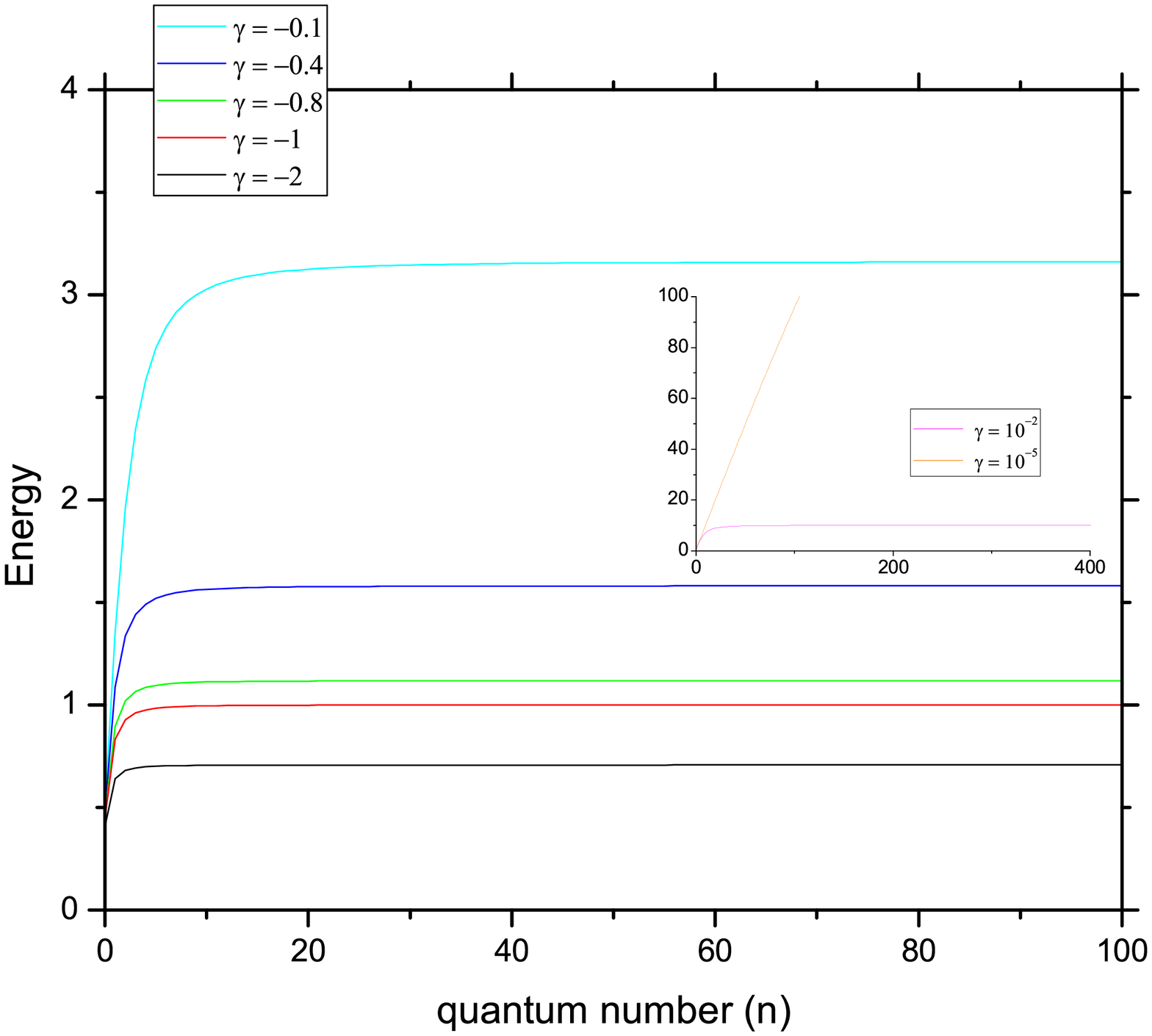}

}

\caption{\label{fig:1}Spectrum of energy $E$ versus quantum number $\left(n\right)$
for both cases.}
\end{figure}

The condition that $\rho_{n}\left(x\right)>0$, leads the the following
implication that $\gamma<0$ for both cases. Now, we are ready to
discuss some interesting results that are not well comments in the
literature. Analytically, the asymptotic limits  for both form of
energies are $\frac{1}{\left|\gamma\right|}$ for the first case,
and $\frac{1}{\sqrt{\left|\gamma\right|}}$ for the second one. These
limits have been reproduced for both cases in Figure . \ref{fig:1}
. Following this figure, some remarks can be made:
\begin{itemize}
\item the modified scalar product is the origin of that the he spectrum
exhibits saturation instead of growing infinitely, 
\item the analytical asymptotic limits are well depicted, 
\item the beginning of the saturation starts from a specific quantum number
$N$,
\item in the limit where $\gamma>-2$, the saturation appears for all values
of quantum number $n$. 
\item finally, when $\gamma$ takes a very small values (in our case $\left|\gamma\right|=10^{-5}$),
we recover the well-known spectrum of energy of one-dimensional Harmonic
oscillator.
\end{itemize}
In what follow, we studied the influence of the modification scalar
product on the thermal properties, especially on the specific heat
curves, and also on the Fisher information parameter $F$ for the
case of the one-dimensional Harmonic oscillator as a model. This study
can be extended for other type of potentials dependent with energy.

\section{The Influence on the thermal properties of 1D Harmonic oscillator}

Lets start with the usual definition of the partition function where
\begin{equation}
Z=\sum_{n=0}^{\infty}e^{-\beta E_{n}},\label{eq:20}
\end{equation}
In our situation, we can rewritten this equation by
\begin{equation}
Z=\underbrace{\sum_{n=0}^{N}e^{-\beta E_{n}}}_{\text{contribution of few levels}}+\underbrace{e^{-\beta\Gamma}}_{\text{term of saturation}},\label{eq:21}
\end{equation}
where $\Gamma=\frac{1}{\left|\gamma\right|}$ for the first choice,
and $\Gamma=\frac{1}{\sqrt{\left|\gamma\right|}}$ for the second.
with the first term is the contribution of all levels until a start
of a saturation behavior, the second is the contribution of saturation
of all levels $n>N$.

Following the new definition of partition function, we can see that
in the limit where $\gamma\rightarrow0$ the specific quantum number
$N\rightarrow\infty$, and so we recover the usual partition function.
Moreover, when $\gamma\rightarrow\infty,$ so $N\rightarrow0$, and
consequently $Z=1$ which means that all levels are in saturation.
\begin{figure}
\subfloat[first case: $\nu=1$]{\includegraphics[scale=0.35]{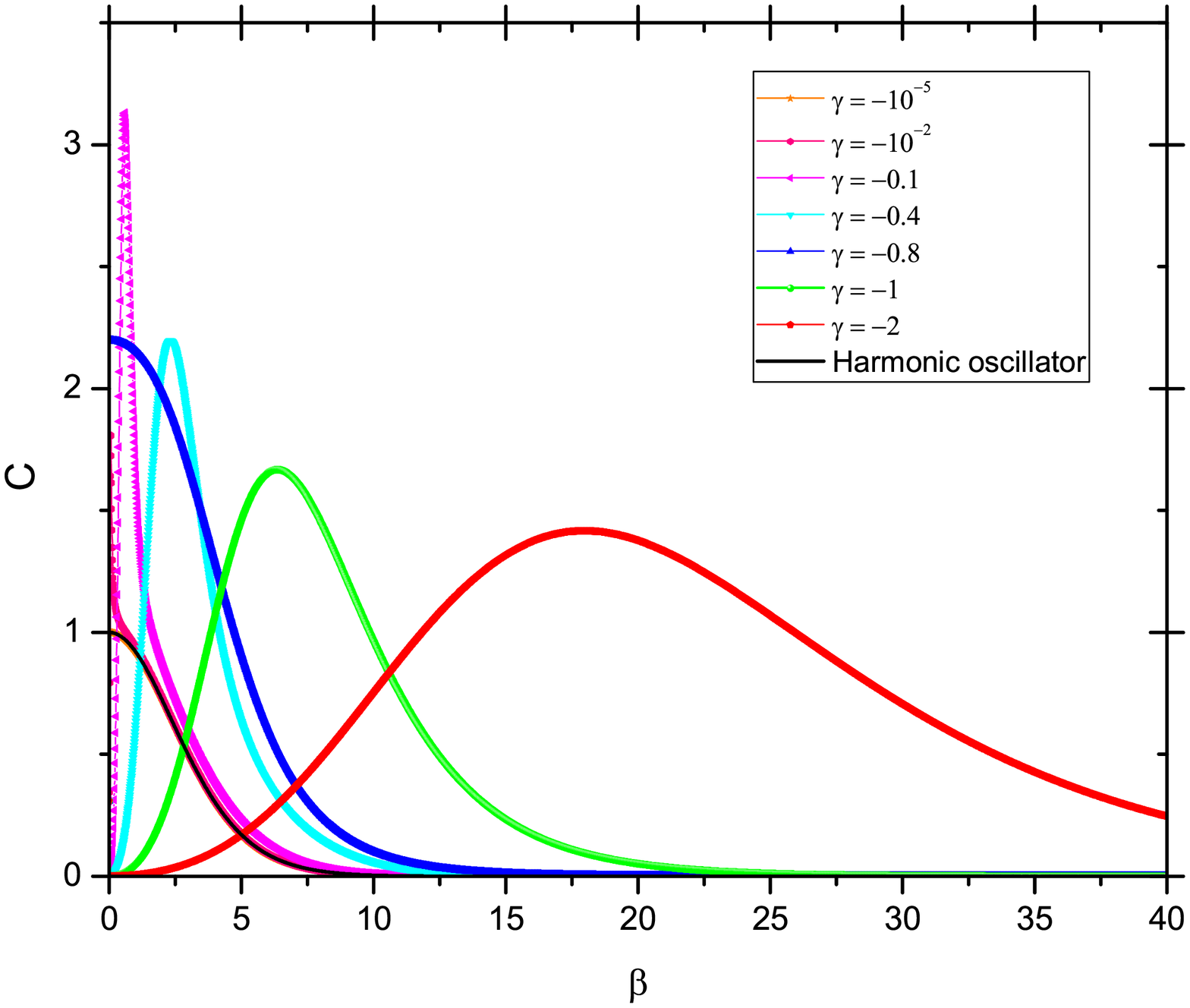}

}\subfloat[second case: $\nu=2$]{\includegraphics[scale=0.35]{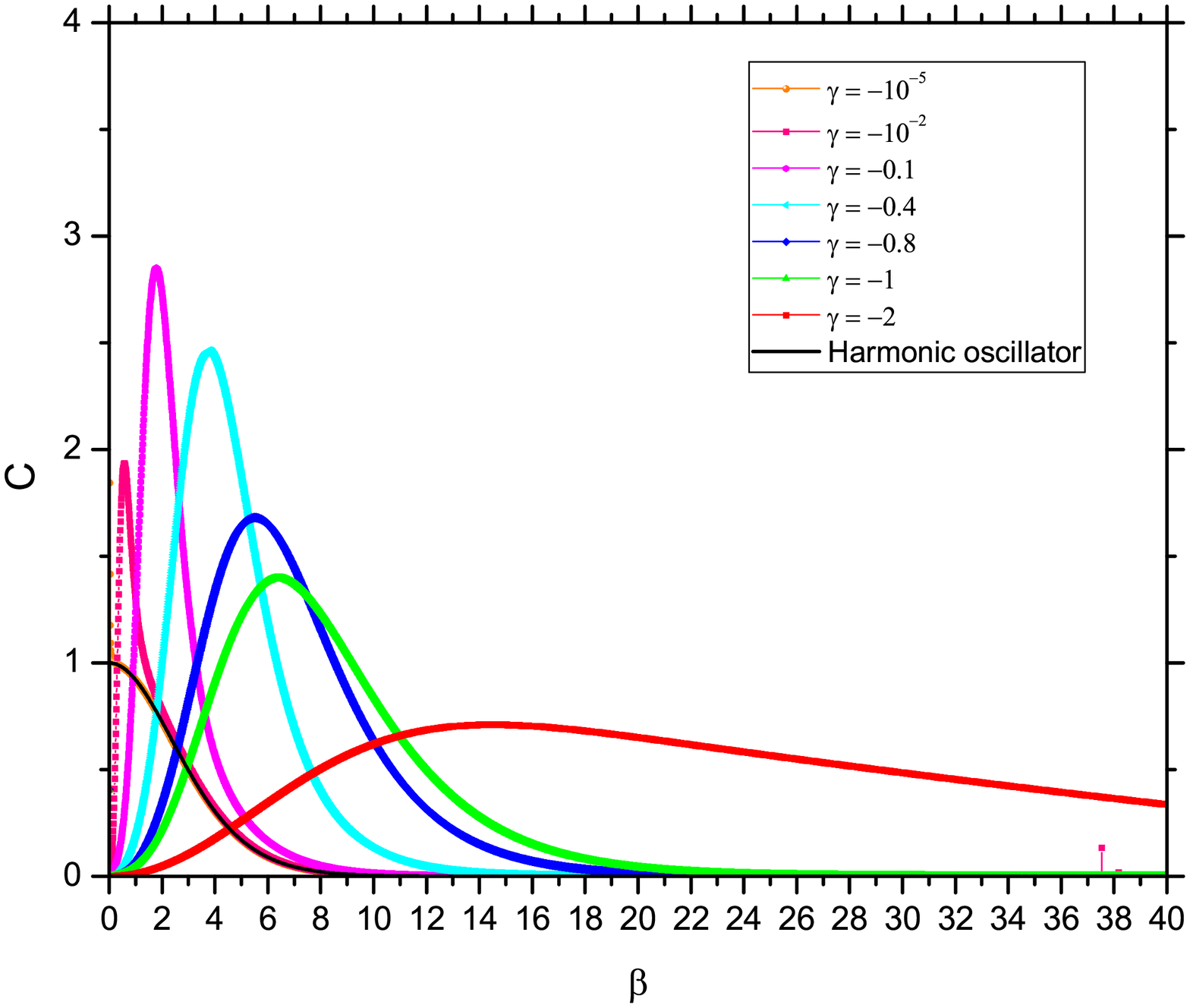}

}

\caption{\label{fig:Cv}Specific heat $C$ versus $\beta$ in both cases}
\end{figure}
 Shown in Figure. \ref{fig:Cv} is the specific heat $C_{v}$ versus
the inverse of temperature $\beta$ for different values of $\gamma$.
Following this figure, one observes that:
\begin{itemize}
\item in the limit where $\gamma$ has a very small values ($\left|\gamma\right|=10^{-2}$
and $\left|\gamma\right|=10^{-5}$), the curve coincide with the one-dimensional
harmonic oscillator in the all range of temperatures. On the other
hand, the curves have the same allure as in the case of the one-dimensional
harmonic oscillator, but tend slowly to zero in a very low-temperatures
($\beta\rightarrow\infty$) for different values of $\gamma$. Now,
in the range of high temperatures ($\beta\rightarrow0$), we have
$C_{v}\rightarrow0$.
\item all curves, except the case of the one-dimensional harmonic oscillator,
exhibit a transition phase between the growth phase and the phase
where the energy shows a saturation.
\end{itemize}
Finally, we can see that the problem of the wave function with energy-dependent
potential leads to the modification of scalar product. This modification
affects the spectrum of energy, and consequently the thermal properties
of the system in question.

\subsubsection{Results and discussions}

The present work is devoted to energy dependent potentials which concerns
the class of potentials having a coupling constant depending linearly
on the energy. In general, the presence of a potential energy dependent
in equation wave induces various significant changes to the usual
rules of mechanical that quantum of which the most obvious is that
the scalar product. the latter is necessary to ensure the conservation
of the norm. According this, we have tested, at first, the effect
of the modified scalar product on the thermal properties of our system
in question through to the spectrum of energy: this modification,
as shown in this section, has imposed a constraint on the $\gamma$
parameter, which leads to the appearance of a phenomenon of saturation
in the diagram of spectrum of energy spectrum (see Fig. \ref{fig:1}).
Then, we have showed that this saturation influence directly the thermal
properties of our $1\text{D}$ Harmonic oscillator, especially on
the curves of specific heat (see Fig. \ref{fig:Cv}). 

We note that the reformulation of the dot product is insufficient
to justify the use of other rules that result in an identical manner
to quantum mechanics usual. In addition, the testing of these formulas
by individual models guarantees not their general validity. This led
Formanek et al \citep{6} to propose a quantum theory with the properties
similar to those of the ordinary non-relativistic quantum theory.
They showed that the wave equation can be transformed into an ordinary
Schrodinger equation with a non-local potential. The clarify this,
equation (\ref{eq:18}) can be rewritten by
\begin{equation}
\rho\left(x\right)=\left[\psi^{*}\left(x\right)\left(1-\frac{\partial V}{\partial E}\right)^{\frac{1}{2}}\right]\left[\left(1-\frac{\partial V}{\partial E}\right)^{\frac{1}{2}}\psi\left(x\right)\right]\equiv\varPsi^{*}\left(x\right)\varPsi\left(x\right),\label{eq:21.1}
\end{equation}
where
\begin{equation}
\varPsi=\left(1-\frac{\partial V}{\partial E}\right)^{\frac{1}{2}}\psi=\mathcal{F}\left(x\right)\psi,\label{eq:21.2}
\end{equation}
and where $\mathcal{F}\left(x\right)=\sqrt{1-\frac{\partial V}{\partial E}}$
is very similar to the Perey Factor \citep{16,17,18,19}: according
to the work of Shimizu et al \citep{19}, we can attribute the wave
function $\varPsi$ to the non-local wave function $\psi_{NL}$, and
the wave function $\psi$ to the equivalent local wave function $\psi_{L}$.
The ratio $\left|\frac{\psi_{NL}}{\psi_{L}}\right|$ forms the Perey
Factor: this factor (more precisely, its deviation from unity), is
in a sense a measure of the nonlocality of the original interaction.
In our case, we see that this factor is greater than unity
\begin{figure}
\includegraphics[scale=0.7]{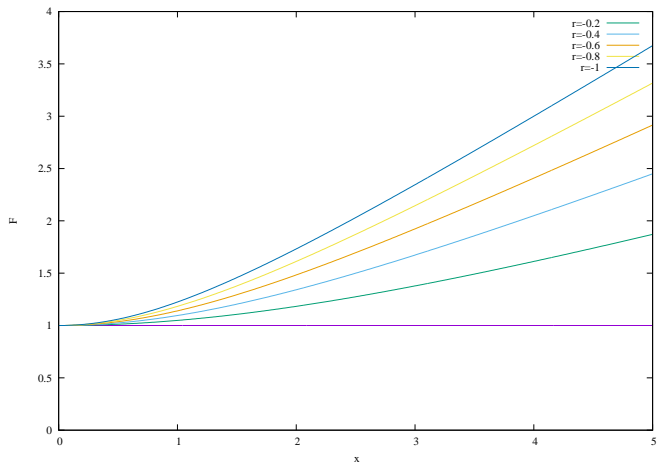}

\caption{Perey factor $\mathcal{F}$ for both cases.}

\end{figure}
. 

Finally, the non-locality and energy dependence of the potential are
known to play an important role in the production of the repulsive
behavior in the baryon-baryon scattering at high energies. It is the
most important feature of the potential based on the quark model employing
the inverse scattering problem \citep{19}.

\section{The influence on the Fisher and Shannon parameters of quantum information}

\subsection{Fisher parameter}

The Fisher information is a quality of an efficient measurement procedure
used for estimating ultimate quantum limits. It was introduced by
Fisher as a measure of intrinsic accuracy in statistical estimation
theory but its basic properties are not completely well known yet,
despite its early origin in 1925. Also, it is the main theoretic tool
of the extreme physical information principle, a general variational
principle which allows one to derive numerous fundamental equations
of physics: Maxwell equations, the Einstein field equations, the Dirac
and Klein-Gordon equations, various laws of statistical physics and
some laws governing nearly incompressible turbulent fluid flows. Fisher
information has been very useful and h)as been applied in different
areas: quantum damped harmonic oscillators, D-dimensional Hydrogenic
systems, Statistical complexity of $\text{H}^{+2}$, position-dependent
mass oscillators, for the case of Tietz-Wei diatomic molecular model,
for the position-dependent mass Schrodinger system, and finally Enhancing
quantum Fisher information by utilizing uncollapsing measurements
\citep{20,21,22,23,24,25,26,27,28,29,30,31,32,33}

\begin{figure}
\subfloat[first case with $\nu=1$.]{\includegraphics[scale=0.3]{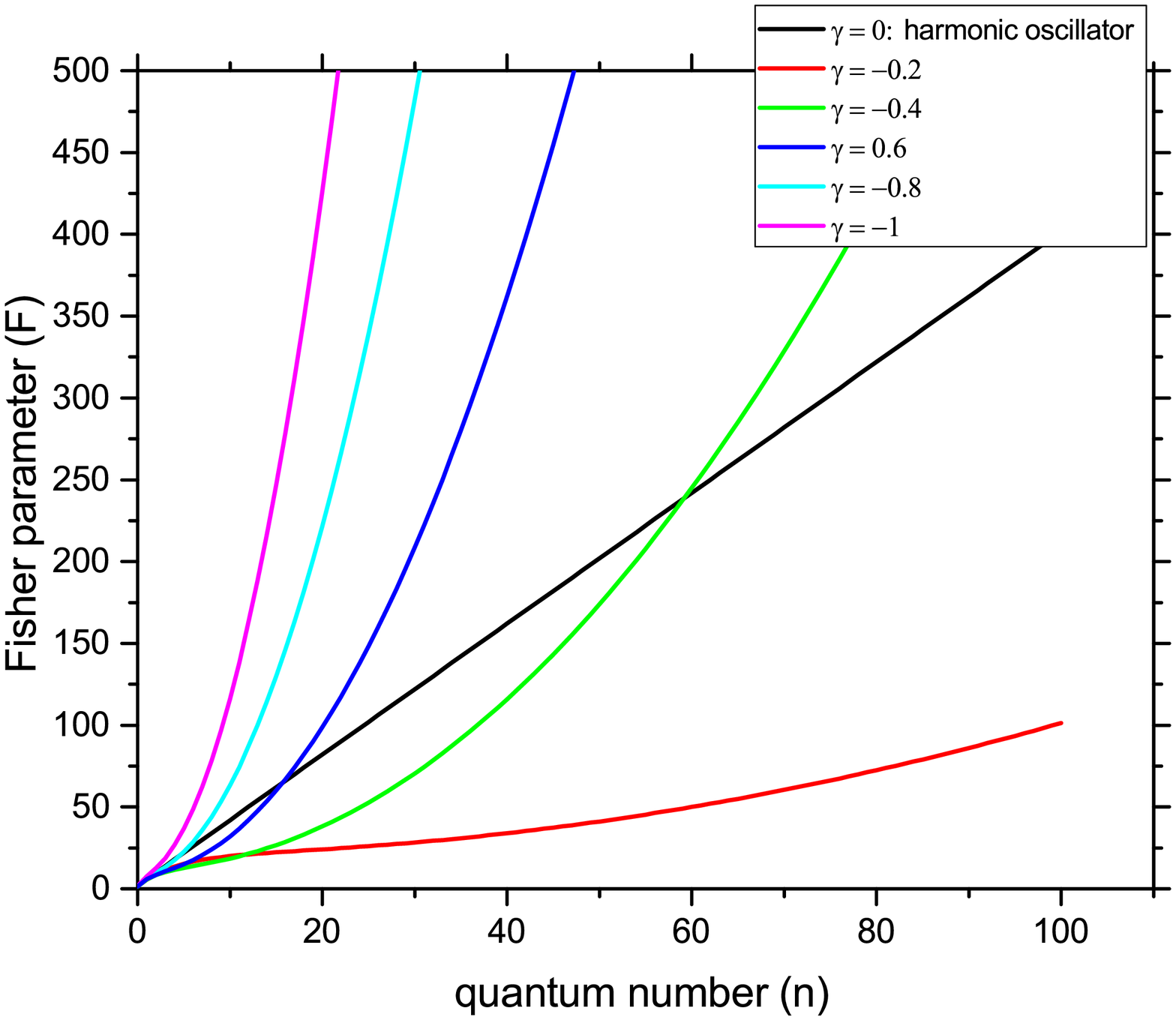}

.}\subfloat[second case with $\nu=2$.]{\includegraphics[scale=0.3]{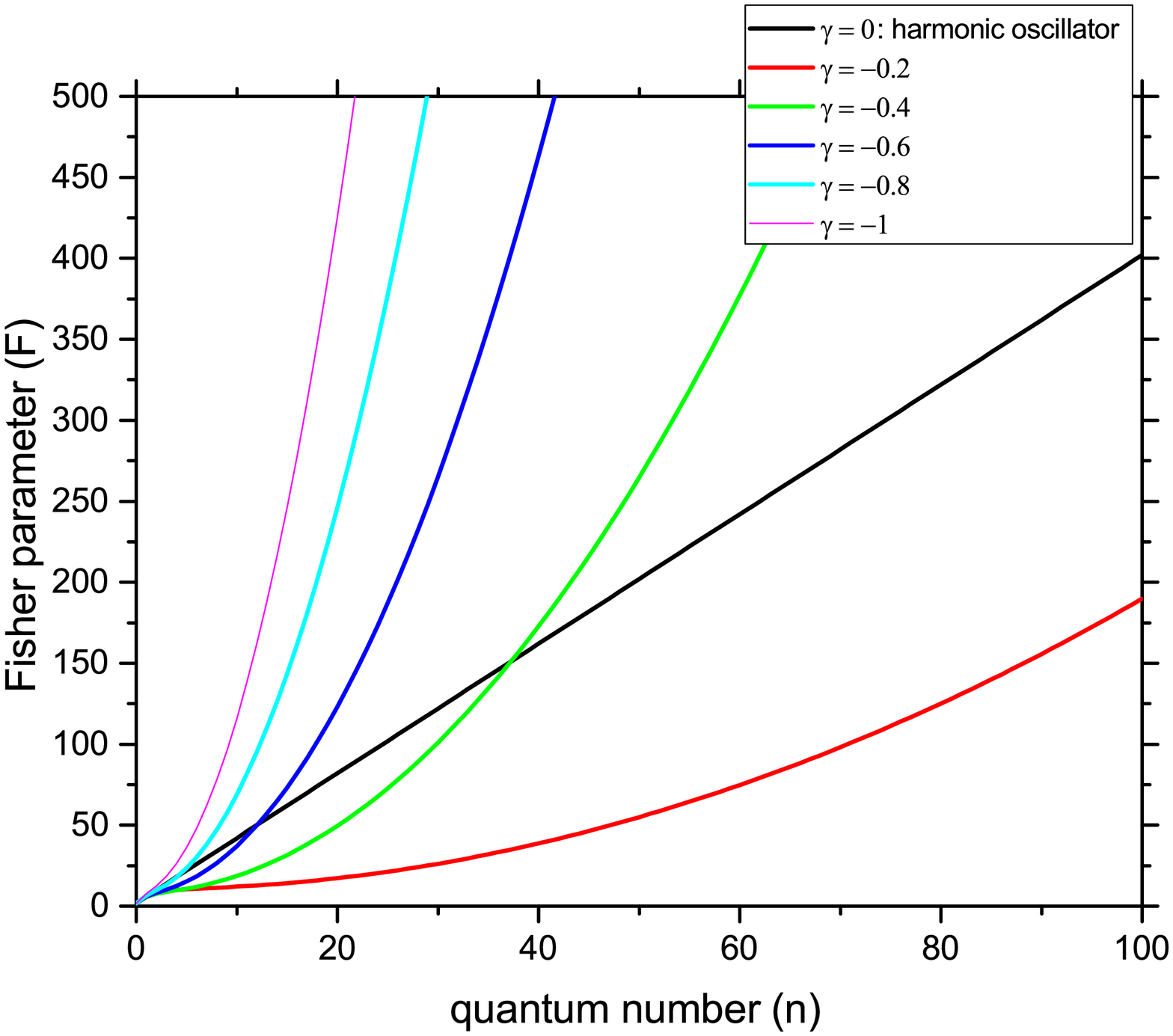}

}

\caption{\label{fig:3}Parameter of Fisher $\left(F\right)$ versus a quantum
number $\left(n\right)$ for both cases. }
\end{figure}

Now, return in our case. The Fisher information of one-dimensional
harmonic oscillator with energy-dependent potential, and with corresponding
probability density $\rho$, is given by
\begin{equation}
F_{x}=\int\rho\left(x\right)\left[\frac{d\text{ln}\rho\left(x\right)}{dx}\right]^{2}dx\geq0,\label{eq:22}
\end{equation}
where
\begin{equation}
\rho_{n}\left(x,\gamma\right)=\left|\psi_{\alpha n}\left(x,\gamma\right)\right|^{2}\left(1-\frac{dV}{dE}\right)=\left|\psi_{\alpha n}\left(x\right)\right|^{2}f\left(x\right),\label{eq:23}
\end{equation}
and where with $f\left(x\right)=1-\frac{\gamma}{2}x^{2}$. By using
the following formula
\begin{equation}
\int_{-\infty}^{+\infty}\left|\psi_{\alpha n}\left(x\right)\right|^{2}\left(1-\frac{dV}{dE}\right)=1,\label{eq:24}
\end{equation}
the constants $C_{\alpha n}$, in both cases, are written by
\begin{equation}
C_{\alpha n}^{2}=\frac{\sqrt{\lambda_{\alpha}}}{2^{n}n!\sqrt{\pi}}\left\{ 1-\frac{\gamma}{4\lambda_{\alpha}}\left(2n+1\right)\right\} ^{-1}.\label{eq:25}
\end{equation}
Thus, the corresponding Fisher information written by
\begin{align}
F_{x}\left(n,\gamma\right) & =\int_{-\infty}^{+\infty}\left(4f\psi'^{2}+4\psi\psi^{'}f+\psi^{2}\frac{f'^{2}}{f}\right)dx,\nonumber \\
 & =\underbrace{\int_{-\infty}^{+\infty}4f\psi'^{2}dx}_{I}+\underbrace{\int_{-\infty}^{+\infty}4\psi\psi^{'}f^{'}dx}_{II}+\underbrace{\int_{-\infty}^{+\infty}\psi^{2}\frac{f'^{2}}{f}dx}_{III},\label{eq:26}
\end{align}
where $\psi'=\frac{d}{dx}\psi$ and $f'=-\gamma x$. After evaluating
the following terms

\begin{equation}
\psi_{\alpha n}'\left(x\right)=C_{\alpha n}e^{-\lambda_{\alpha}\frac{x^{2}}{2}}\left\{ -\lambda_{\alpha}xH_{n}\left(\sqrt{\lambda_{\alpha}}x\right)+2n\sqrt{\lambda_{\alpha}}H_{n-1}\left(\sqrt{\lambda_{\alpha}}x\right)\right\} ,\label{eq:27}
\end{equation}
\begin{align}
4f\psi_{\alpha n}'^{2} & =C_{\alpha n}^{2}e^{-\lambda_{\alpha}x^{2}}\left\{ 4\lambda_{\alpha}^{2}x^{2}H_{n}\left(\sqrt{\lambda_{\alpha}}x\right)^{2}+16n^{2}\lambda_{\alpha}H_{n-1}\left(\sqrt{\lambda_{\alpha}}x\right)^{2}-16n\lambda_{\alpha}^{\frac{3}{2}}xH_{n}\left(\sqrt{\lambda_{\alpha}}x\right)H_{n-1}\left(\sqrt{\lambda_{\alpha}}x\right)\right\} ,\nonumber \\
 & +C_{\alpha n}^{2}e^{-\lambda_{\alpha}x^{2}}\left\{ -2\gamma\lambda_{\alpha}^{2}x^{4}H_{n}\left(\sqrt{\lambda_{\alpha}}x\right)^{2}-8n^{2}\gamma\lambda_{\alpha}x^{2}H_{n-1}\left(\sqrt{\lambda_{\alpha}}x\right)^{2}+8n\gamma\lambda_{\alpha}^{\frac{3}{2}}x^{3}H_{n}\left(\sqrt{\lambda_{\alpha}}x\right)H_{n-1}\left(\sqrt{\lambda_{\alpha}}x\right)\right\} \label{eq:28}
\end{align}
\begin{equation}
4\psi\psi^{'}f^{'}=\gamma C_{\alpha n}^{2}e^{-\lambda_{\alpha}x^{2}}\left\{ 4\lambda_{\alpha}x^{2}H_{n}\left(\sqrt{\lambda_{\alpha}}x\right)^{2}-8n\sqrt{\lambda_{\alpha}}xH_{n}\left(\sqrt{\lambda_{\alpha}}x\right)H_{n-1}\left(\sqrt{\lambda_{\alpha}}x\right)\right\} ,\label{eq:29}
\end{equation}
\begin{equation}
\psi^{2}\frac{\left(f'\right)^{2}}{f}=\frac{\gamma^{2}C_{\alpha n}^{2}e^{-\lambda_{\alpha}x^{2}}x^{2}H_{n}^{2}\left(\sqrt{\lambda_{\alpha}}x\right)}{1-\frac{\gamma}{2}x^{2}},\label{eq:30}
\end{equation}
we obtain that 
\begin{equation}
I=\frac{4\lambda_{\alpha}\left(n+\frac{1}{2}\right)-4n\gamma-\frac{\gamma}{2}\left[\left(2n+1\right)^{2}+2\right]}{1-\frac{\gamma}{4\lambda_{\alpha}}\left(2n+1\right)},\label{eq:31}
\end{equation}
the second
\begin{equation}
II=\frac{2\gamma}{1-\frac{\gamma}{4\lambda_{\alpha}}\left(2n+1\right)},\label{eq:32}
\end{equation}
\begin{align}
III & =\gamma^{2}C_{\alpha n}^{2}\intop_{-\infty}^{+\infty}\frac{e^{-\lambda_{\alpha}x^{2}}x^{2}H_{n}^{2}\left(\sqrt{\lambda_{\alpha}}x\right)}{1-\frac{\gamma}{2}x^{2}}dx=\frac{\gamma^{2}C_{\alpha n}^{2}}{\lambda_{\alpha}^{\frac{3}{2}}}\intop_{-\infty}^{+\infty}\frac{e^{-y^{2}}y^{2}H_{n}^{2}\left(y\right)}{1-\frac{\gamma}{2\lambda}y^{2}}dy.\nonumber \\
 & \simeq\frac{\gamma^{2}C_{\alpha n}^{2}}{\lambda_{\alpha}^{\frac{3}{2}}}\left\{ \intop_{-\infty}^{+\infty}e^{-y^{2}}y^{2}H_{n}^{2}\left(y\right)dy-\frac{\gamma}{2\lambda}\intop_{-\infty}^{+\infty}e^{-y^{2}}y^{4}H_{n}^{2}\left(y\right)dy\right\} \label{eq:33}
\end{align}
where $y=\sqrt{\lambda_{\alpha}}x$, and where 
\begin{equation}
\frac{1}{1-\frac{\gamma}{2\lambda}y^{2}}=\sum_{n=0}^{\infty}\left(-1\right)^{n}\left(\sqrt{\frac{\left|\gamma\right|}{2\lambda}}y\right)^{2n}\simeq1-\frac{\gamma}{2\lambda}y^{2}+\ldots.\label{eq:34}
\end{equation}
So, we have
\begin{equation}
III=\frac{\frac{\gamma^{2}\left(n+\frac{1}{2}\right)}{\lambda_{\alpha}}-\frac{\gamma^{3}\left[\left(2n+1\right)^{2}+2\right]}{8\lambda_{\alpha}^{2}}}{1-\frac{\gamma}{4\lambda_{\alpha}}\left(2n+1\right)}.\label{eq:35}
\end{equation}
Following these results, the final form of Fisher parameter is
\begin{align}
F_{\alpha}\left(n,\gamma\right) & =\left\{ 1-\frac{\gamma}{4\lambda_{\alpha}}\left(2n+1\right)\right\} ^{-1}\times\nonumber \\
 & \left\{ 4\lambda_{\alpha}\left(n+\frac{1}{2}\right)-4n\gamma-\frac{\gamma}{2}\left[\left(2n+1\right)^{2}+2\right]+2\gamma+\frac{\gamma^{2}\left(n+\frac{1}{2}\right)}{\lambda_{\alpha}}-\frac{\gamma^{3}\left[\left(2n+1\right)^{2}+2\right]}{8\lambda_{\alpha}^{2}}\right\} .\label{eq:36}
\end{align}
This formula are plotted in the Figure. \ref{fig:3}. This figure
showed the parameter of Fisher versus a quantum number for different
values of $\gamma$ in both cases: the case of harmonic oscillator
is also inserted for comparison. We observe that the Fisher parameter
$F_{\alpha}\left(\alpha=1,2\right)$ increases with $n$ for different
values of $\gamma$. This increase indicates that $\rho_{n}\left(x,\gamma\right)$
becomes more and more localized, increasing the accuracy in predicting
the localization of the particle \citep{32}. The comparison with
the case of one-dimensional Harmonic oscillator ($\gamma=0$) showed
that for the values of $\left|\gamma\right|<0.3$, the curves of the
Fisher parameter are below to the case of 1D harmonic oscillator.
On the other hand, and from the values of $\left|\gamma\right|>0.4$
, these curves move above this last. 

\begin{figure}
\subfloat[first case where $\nu=1$.]{\includegraphics[scale=0.3]{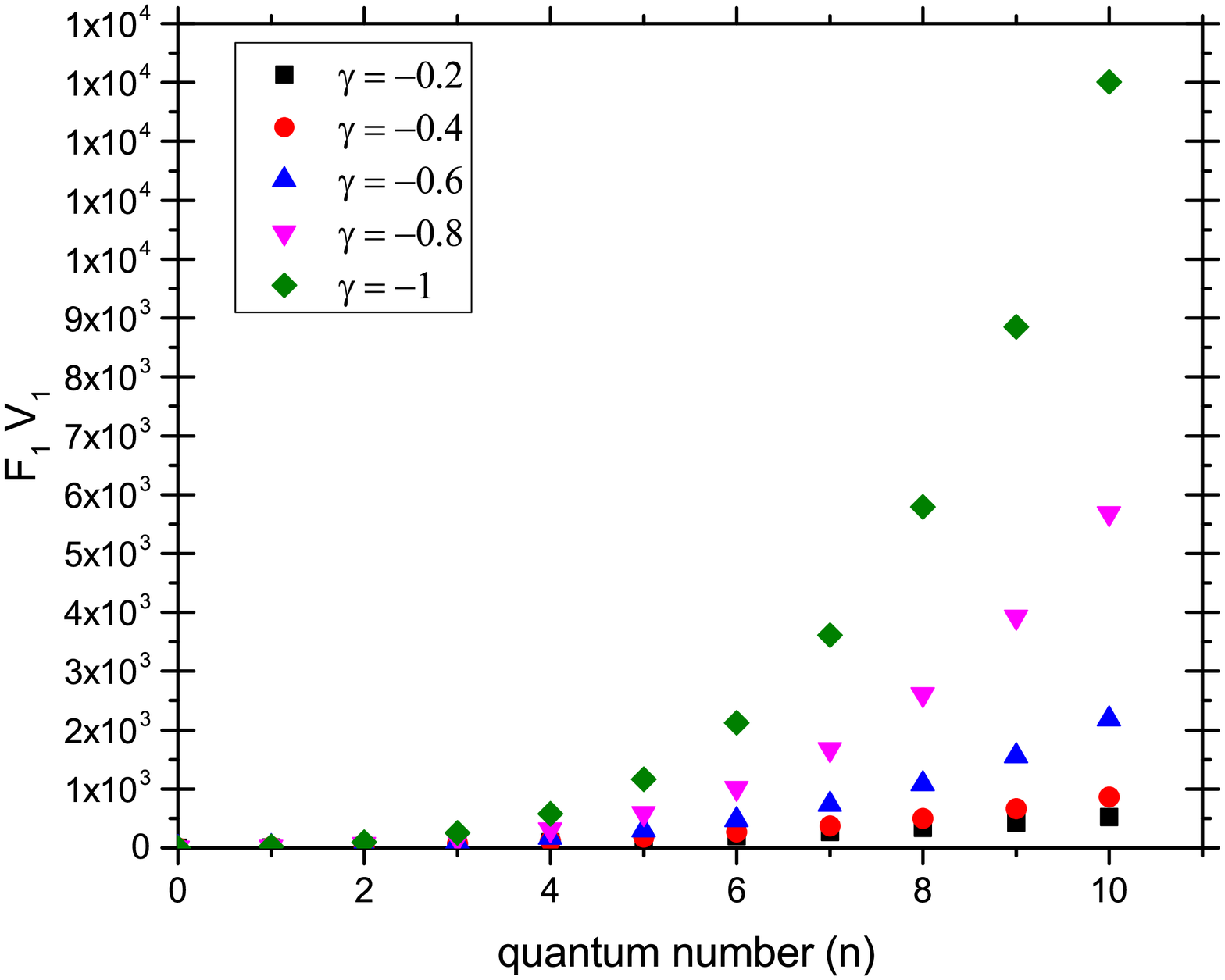}

}\subfloat[second case where $\nu=2$.]{\includegraphics[scale=0.3]{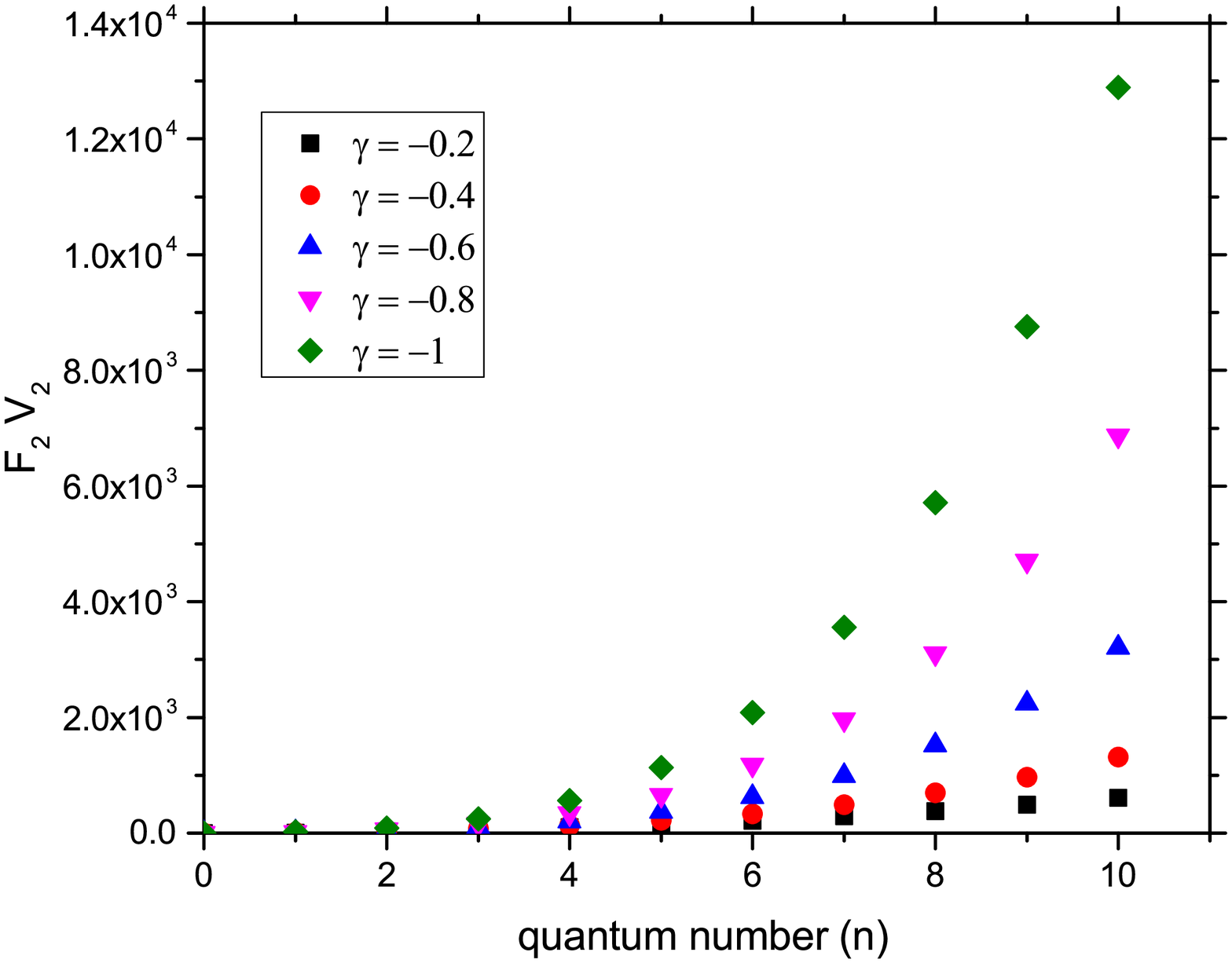}

}

\caption{\label{fig:4}The Cramer-Rao uncertainty relation versus quantum number
$n$ for different values of $\gamma$.}
\end{figure}

Now, in order to justify our equation for the parameter of Fisher,
the well-known Cramer-Rao uncertainty relation 
\begin{equation}
F_{\alpha}\cdot V_{\alpha}\geq1,\label{eq:37}
\end{equation}
with $V_{\alpha}=\left\langle x^{2}\right\rangle _{\alpha}-\left\langle x\right\rangle _{\alpha}^{2}$
must be fulfilled. This parameter is an inequality which involves
fisher information and variance. The variance $V_{\alpha}$ requires
the knowledge of both parameters $\left\langle x^{2}\right\rangle _{\alpha}$
and $\left\langle x\right\rangle _{\alpha}^{2}$. In our case, the
modification of the scalar product affects leads to the following
equations:
\begin{equation}
\left\langle x\right\rangle _{\alpha}=\intop_{-\infty}^{+\infty}x\psi_{n}^{2}\left(x\right)f\left(x\right)dx=0,\label{eq:38}
\end{equation}
and
\begin{equation}
\left\langle x^{2}\right\rangle _{\alpha}=\intop_{-\infty}^{+\infty}x^{2}\psi_{n}^{2}\left(x\right)f\left(x\right)dx=\frac{2n+1}{2\lambda_{n}}\left\{ 1-\frac{\gamma}{4\lambda_{n}}\left(2n+1\right)\right\} ^{-1}\left\{ 1-\frac{\gamma}{4\lambda_{n}}\frac{\left[\left(2n+1\right)^{2}+2\right]}{\left(2n+1\right)}\right\} .\label{eq:39}
\end{equation}
The uncertainty relation of Cramer-Rao is depicted in Figure. \ref{fig:4}.
From this Figure, we can see that this parameter increases with $\gamma$,
and the Eq. (\ref{eq:37}) is well recovered.

\subsection{Shannon entropy}

Entropic measures provide analytic tools to help us to understand
correlations in quantum systems. Shannon has introduced entropy to
measure the uncertainty. Now, it has become a universal concept in
statistical physics. The Shannon entropy has finding applications
in several branches of physics because of its possible applications
in a wide range of area such as to the nuclei, for the Morse and the
Poschl-Teller potentials, in the relationship between the densities
of Shannon entropy and Fisher information for atoms and molecules,
in the study of the the position and momentum information-theoretic
measures of a D-dimensional particle in a box (see Ref. (\citep{35})
and references therein).

\begin{figure}
\subfloat[$\nu=1$]{\includegraphics[scale=0.85]{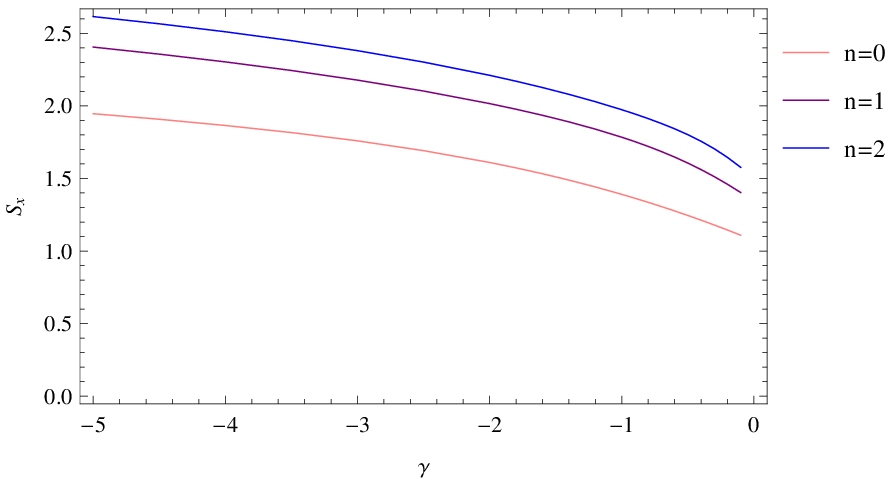}

}\subfloat[$\nu=2$]{\includegraphics[scale=0.85]{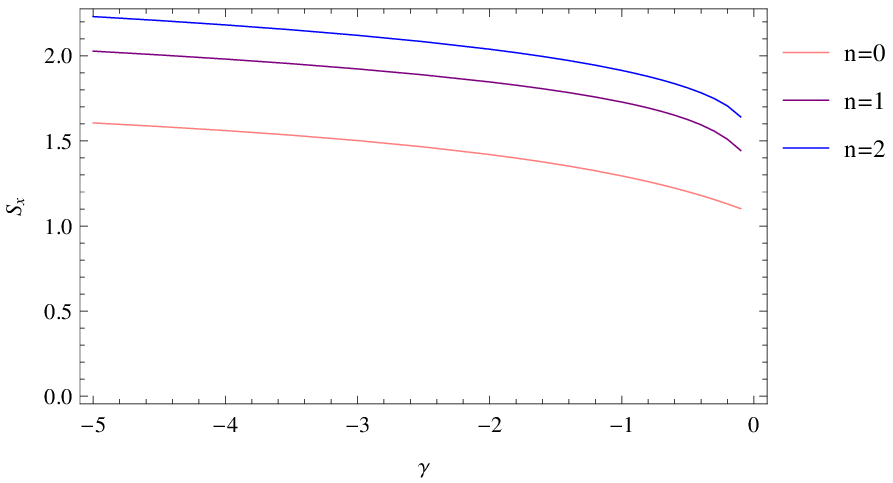}

}

\caption{\label{fig:5}Behavior of $S_{x}$ versus $\gamma$ varying for different
values of $n$}
\end{figure}

\begin{figure}
\subfloat[$\nu=1$]{\includegraphics[scale=0.75]{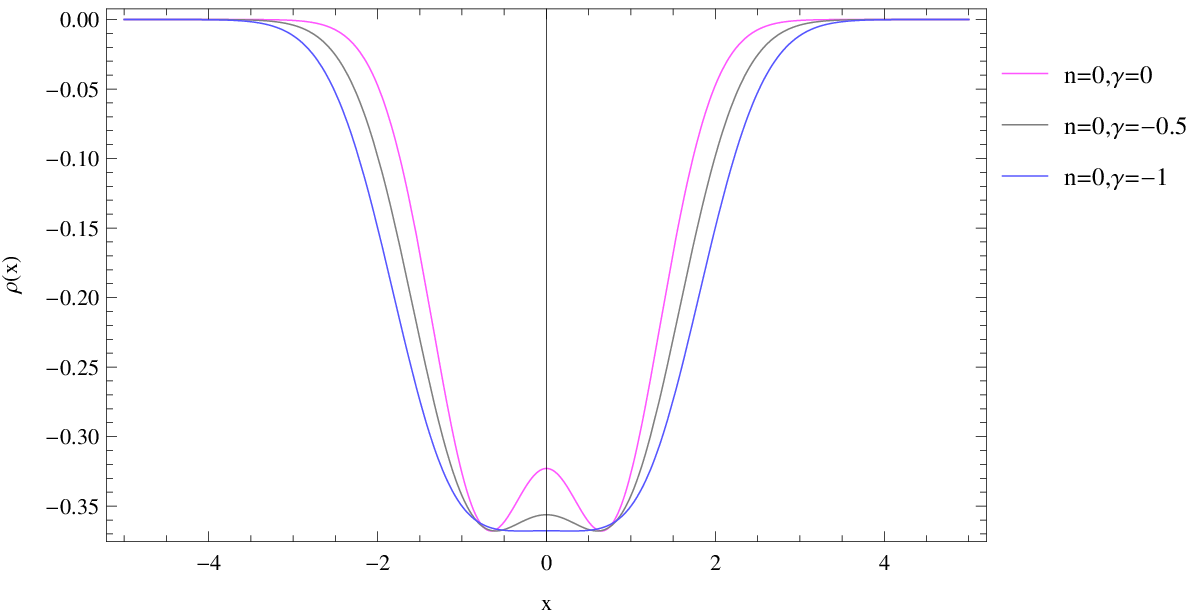}

}\subfloat[$\nu=2$]{\includegraphics[scale=0.75]{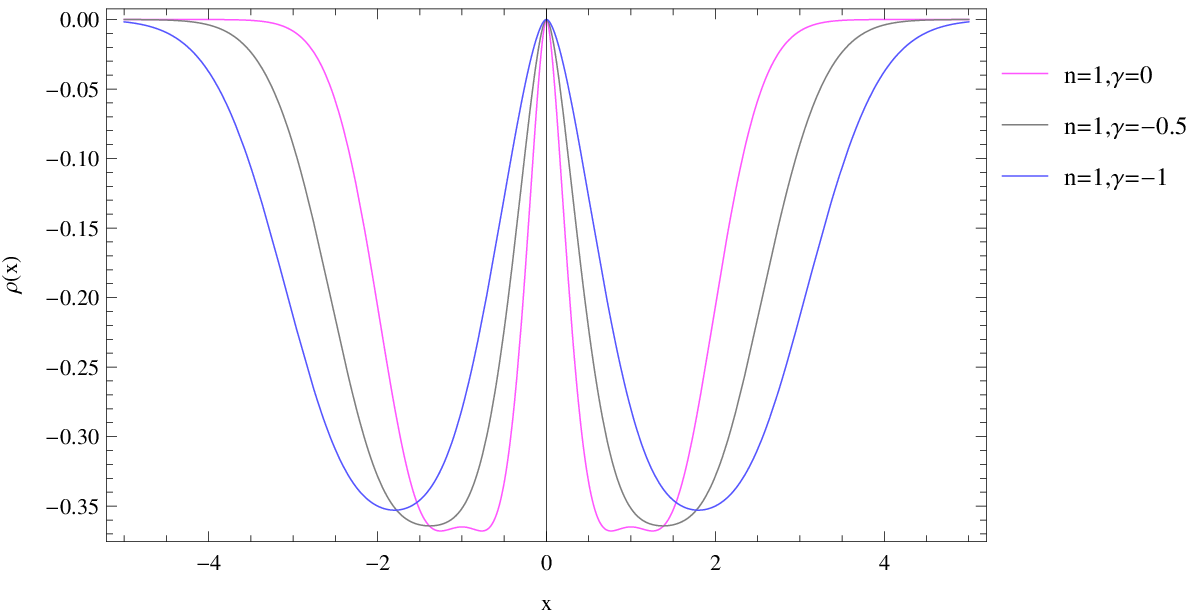}

}

\caption{\label{fig:6}Behavior of $\rho(x)$ versus the variable $x$ }

\end{figure}

The position space information entropies for the one-dimensional can
be calculated by using \citep{36,37,38,39,40,41}
\begin{equation}
S_{x}=-\int\left|\psi\left(x\right)\right|^{2}\ln\left|\psi\left(x\right)\right|^{2}dx,\label{eq:40}
\end{equation}
with $\psi(w)$ being the normalized eigenfunction. In our case, the
above equation becomes
\begin{equation}
S_{x}=-\int\rho_{n}\left(x,\gamma\right)\ln\rho_{n}\left(x,\gamma\right)dx,\label{eq:41}
\end{equation}
with $\rho_{n}\left(x,\gamma\right)$ is defined by the equation (\ref{eq:23}).
In general, explicit derivations of the information entropy are quite
difficult. In particular, the derivation of analytical expression
for the $S_{x}$ is almost impossible. We represent the position information
entropy densities respectively by
\begin{equation}
\rho_{S_{x}}=\rho_{n}\left(x,\gamma\right)\ln\rho_{n}\left(x,\gamma\right).\label{eq:42}
\end{equation}
We have plotted the results of this section in some figures. In Fig.
\ref{fig:5} , we consider $\hbar=1,m=1,\omega=1,\nu=1,2$ for difference
$n$ , where we plotted the Shannon information entropy $S_{x}$ versus
$\gamma$. 
We also plotted these entropy densities for$\hbar=1,m=1,\omega=1,\nu=1,2$
for difference $n$ and $\gamma$ in Fig. \ref{fig:6}
.

\subsection{Results and discussions}

Fisher information is an important concept in quantum estimation and
quantum information theory. Its dependence on the probability of density
function $\rho$, allows us to test the influence of the modified
scalar product, in the case of the problems of the wave function with
energy-dependent potentials, on this parameter. The correspondence
between a problem of wave function with an ordinary Schrodinger equation
with a non-local potential permits us to consider the dependence of
this parameter with non-locality of the potential via a coupling constant
depending linearly on the energy,i.e., the parameter $\gamma$: In
this section we shown that this factor, in both cases, depends strongly
with $\gamma$: thus, the non-locality of the potential has a direct
influence on the parameter of Fisher. Moreover, we have used the well-known
Cramer-Rao uncertainty relation in order to justify our results concerning
the Fisher parameter.

In the same way, this influence is well established for the case of
Shannon entropy: the information entropies of the ground and the excited
states are versus $\gamma$ is shown in Figs. \ref{fig:5}. Through
this parameter, the influence of the modified scalar product on $S_{x}$
is very clear for both cases in consideration. The Fig. \ref{fig:6}
illustrates the position information entropy densities $\rho_{s}(x)=\left|\psi_{\alpha n}\left(x,\gamma\right)\right|^{2}\left(1-\frac{dV}{dE}\right)$
. They play a role similar to that of the probability density $\rho(x)=\left|\psi_{\alpha n}\left(x,0\right)\right|^{2}$
in quantum mechanics.

\section{Conclusion}

The present work is devoted to energy dependent potentials. We have
considered the cases of the one-dimensional Harmonic oscillator with
energy dependence. This case leads to a coherent theory. As a first
result, we show that the energy dependence affects essentially the
eigensolutions. Especially, we observe a saturation on the curves
of the spectrum of energy. Also, the presence of the energy dependent
potential in a wave equation leads to the modification of the scalar
product, necessary to ensure the conservation of the norm. So, we
have shown that this modification affects directly, (i) the thermal
properties of our systems, and (ii) the form of the parameters of
Fisher and Shannon. The uncertainty relation of Cramer-Rao is well-established,
which means that our calculations on the Fisher parameter are correct.

\end{document}